\documentclass[iop,twocolumn]{emulateapj}		

\usepackage[]{natbib}
\usepackage{graphicx}	
\usepackage{amsmath}
\usepackage[T1]{fontenc}		
\usepackage{color}			
\usepackage[normalem]{ulem}
\usepackage[protrusion=true,expansion=false]{microtype} 

\newcounter{species}

\newcommand{\kms}{km~s$^{-1}$}

\newcommand{\Msun}{\ensuremath{M_{\scriptscriptstyle \odot}}}

\newcommand{\OIIIw}{[O$\,\textsc{iii}]$~$\lambda$5007}
\newcommand{\OIIIdblt}{[O$\,\textsc{iii}]$~$\lambda\lambda$4959, 5007}

\newcommand{\OIII}{[O\,{\sc iii}]}

\newcommand{\COIO}{CO\,$(1\rightarrow0)$}

\def\lsim{\lower0.3em\hbox{$\,\buildrel <\over\sim\,$}}
\def\gsim{\lower0.3em\hbox{$\,\buildrel >\over\sim\,$}}

\def\kms{\,km~s$^{-1}$}      

\def\deg{\hbox{$^\circ$}}

\def\lesssim{\mathrel{\hbox{\rlap{\hbox{%
 \lower4pt\hbox{$\sim$}}}\hbox{$<$}}}}
\def\gtrsim{\mathrel{\hbox{\rlap{\hbox{%
 \lower4pt\hbox{$\sim$}}}\hbox{$>$}}}}

\def\arcmin{\hbox{$^\prime$}}
\def\arcsec{\hbox{$^{\prime\prime}$}}

\def\fs{\hbox{$.\!\!^{\rm s}$}}

\def\farcs{\hbox{$.\!\!^{\prime\prime}$}}

\def\fhnd{\hbox{$^{\rm h}$}}

\def\fdgnd{\hbox{$^\circ$}}

\shorttitle{Jet-driven molecular outflow in PG~1700$+$518?}
\shortauthors{Runnoe et al.}

\begin{document}		
\title{Does the compact radio jet in PG~1700+518 drive a molecular outflow?}

\author{
Jessie C. Runnoe\altaffilmark{1}, Kayhan G{\"u}ltekin\altaffilmark{1}, David S. N. Rupke\altaffilmark{2,3}}
\altaffiltext{1}{Department of Astronomy, University of Michigan, 1085 S. University Ave., Ann Arbor, MI 48109}
\altaffiltext{2}{Department of Physics, Rhodes College, Memphis, TN 38112, USA}
\altaffiltext{3}{Research School of Astronomy and Astrophysics, The Australian National University, Canberra, ACT 2611, Australia}



\begin{abstract}
Radio jets play an important role in quasar feedback, but direct observations showing how the jets interact with the multi-phase interstellar medium of galaxy disks are few and far between.  In this work, we provide new millimeter interferometric observations of PG~1700+518 in order to investigate the effect of its radio jet on the surrounding molecular gas.  PG~1700 is a radio-quiet, low-ionization broad absorption line quasar whose host galaxy has a nearby interacting companion.  On sub-kiloparsec scales, the ionized gas is driven to high velocities by a compact radio jet that is identified by radio interferometry.  We present observations from the NOrthern Extended Millimeter Array (NOEMA) interferometer with a 3\farcs8 (16~kpc) synthesized beam where we detect the \COIO\ emission line at $30\sigma$ significance with a total flux of $3.12\pm0.02$~Jy~\kms\ and a typical velocity dispersion of $125\pm5$\kms.  Despite the outflow in ionized gas, we find no concrete evidence that the CO gas is being affected by the radio jet on size scales of a kiloparsec or more.  However, a $\sim1$\arcsec\ drift in the spatial centroid of the CO emission as a function of velocity across the emission line and the compact nature of the jet hint that higher spatial resolution observations may reveal a signal of interaction between the jet and molecular gas.
\end{abstract}

\section{Introduction}
Early analytical work \citep{silk98,fabian99} predicts that an active galactic nucleus (AGN) can heat or even expel the gas in its host galaxy, thus halting both star formation and accretion onto the central supermassive black hole  \citep{king03} if the mechanical pressure, radiation pressure, or thermal heating from the AGN can couple to gas in the host galaxy \citep{faucher-giguere12}.  This kind of AGN feedback may be the natural result of galaxy merger scenarios \citep{dimatteo05,springel05} and has the potential to explain an increasing number of observations.  Among these, are the tight empirical relationships between the properties of supermassive black holes and their host galaxies \citep[e.g.,][]{magorrian98,gultekin09}, the roughly constant ratio between star formation and black hole growth with cosmic time \citep{kauffmann03,lamassa12a,heckman14,gurkan15}, and the enrichment of the Galactic halo and circumgalactic medium with metals \citep{veilleux13b}.

However, there is disagreement as to what extent different kinds of AGN activity actually regulate galaxy growth.  In particular, it is not clear whether or when the central engine and its host are coupled through the radiative (quasar-mode) or the kinematic (radio-mode) impact of the AGN \citep[for a review of AGN feedback modes, see][]{fabian12}.  In this work, we investigate  the impact of AGN radio jets, which do seem to be capable of affecting the host galaxy on large scales \citep[at least in some gas phases,][]{guillard12}.  Notably, semi-analytic models which include prescriptions for radio-mode feedback \citep[e.g.,][]{bower06,croton06} are able to better reproduce some empirical findings, including the deficit at the bright end of the galaxy luminosity function.  By pumping energy into the hot, tenuous circumgalactic medium, the AGN prevents the infall of cool gas, halts star formation, and thus limits galaxy growth.

The above implementations of radio-mode feedback concern primarily massive elliptical galaxies at the centers of clusters \citep[although see e.g.,][for the full impact of radio-mode feedback]{somerville08} and do not capture the local effect of the central engine on the interstellar medium (ISM) of less massive disk galaxies, which have very different conditions.  The observed absence of correlations between feedback-related galaxy and jet properties suggests that the heating and turbulence imparted to the host ISM via radio-mode feedback is not the sole driver suppressing star formation in less massive systems \citep{lanz16}.  The lack of correlations between jet and galaxy properties doesn't preclude the possibility that in disk galaxies, the radio jets may interact directly with the ISM of the host and regulate galaxy growth by expelling the ingredients for forming stars in high-velocity outflows.  

Three dimensional hydrodynamic simulations of relativistic jets interacting with a multi-phase disk, which have dense clumps embedded in a less dense medium, suggest that jets may ultimately suppress star formation via heating and ablation of the dense clumps \citep{wagner11,wagner12,gaibler12}.  Extending the work of \citet{wagner11} and \citet{wagner12}, \citet{mukherjee16} initializes the host ISM in a realistic gravitational potential with turbulence consistent with high-redshift galaxies.  In these simulations the radio plasma can have an irregular morphology; the jet takes the path of least resistance through the clumpy ISM and can be redirected relatively easily.  The energy bubble driven by the jet creates multiple phases in the ISM, and radial outflows are accelerated at the shocked surface of the bubble.  The main result of this work is that low-power jets ($P_{jet} \lesssim 10^{43}$~erg~s$^{-1}$) can significantly impact star formation in their host galaxies over a large kiloparsec-scale volume.  These jets lack the momentum to punch through the dense nuclear gas and the resulting energy bubble imparts turbulence to the ISM as it spreads laterally.  High-power jets ($P_{jet}\gtrsim 10^{45}$~erg~s$^{-1}$) can more efficiently accelerate high-velocity outflows ($v\lesssim500$\kms), but do not impact their hosts over as large of a volume.
	
Observational evidence for the direct impact of AGN jets on the ISM of the host galaxy has proven rare, but not completely elusive \citep{morganti05,sakamoto14,garcia-burillo14,aalto16}.  There are two noteworthy examples of jet-driven cold molecular outflows.  IC~5063, a type 2 Seyfert nucleus hosted in an early-type galaxy, may be the best example of a radio jet driving a fast molecular outflow.  Observations paint a picture where the molecular gas, like the neutral gas, has been accelerated as a result of interactions between expanding radio lobes and the ISM in the disk of the host galaxy \citep{tadhunter14,dasyra16,oosterloo17}.  NGC~1266, which has a suppressed star formation rate \citep{alatalo15a}, also has a small radio jet \citep{nyland13} which drives a molecular outflow \citep{alatalo11}.

Nearby AGN like PG~1700+518 \citep[hereafter PG~1700,][]{schmidt83}, which have signs of jet-driven outflows in some gas phases and extensive multi-wavelength data, present excellent opportunities to investigate how jets interact with host galaxies in practice.  PG~1700 is an intermediate-redshift ($z=0.2902$) type 1 Seyfert nucleus with high- and low-ionziation broad absorption lines \citep{turnshek85,wampler85,pettini85,young07} hosted by a galaxy undergoing a merger with a nearby companion \citep{stockton98,hines99}.  It is X-ray weak, with a steep spectrum indicative of either strong absorption or a reflection-dominated component \citep{ballo11}.  

A wealth of radio observations reveal that although the nucleus is radio quiet, it is not radio silent \citep{barvainis89b,barvainis96}.  At low frequencies it is unresolved by the Very Large Array (VLA).  The 1.4~GHz logarithmic radio luminosity adjusted for our cosmology is 24.65~W~Hz$^{-1}$ \citep{kukula98}, well below the knee of the radio luminosity function \citep{condon13} and at the border between the Fanaroff-Riley (FR) I and II classifications.  Using the FR~I relation of \citet{godfrey16}, this corresponds to a jet power of order $P_{jet}\sim10^{45-46}$~erg~s$^{-1}$.  At higher frequencies ($>5$~GHz) the source has two components separated by $\sim$1\arcsec\ \citep{hutchings92,kellerman94,kukula98}.  High spatial resolution interferometry with the European Very Long Baseline Interferometry Network (EVN) reveals a radio core and two-sided jet in the western of the two radio components \citep{yang12}.  

PG~1700 has a one-sided extended narrow-line region \citep[NLR,][]{husemann13} and an ionized outflow along the jet axis traced by \OIIIw\ at high velocities \citep{rupke17}.  \citet{evans09} present the first millimeter observations of the \COIO\ from the Institute de Radioastronomie Millim{\'e}trique (IRAM) 30m, demonstrating that with $M(H_2)\sim6\times10^{10}$~\Msun, PG~1700 has one of the most molecular gas-rich host galaxies of the Palomar-Green quasars.  The single-dish observations show the presence of molecular gas at velocities in excess of 100\kms\ from systemic, but with a 26\arcsec\ beam they lack the spatial resolution to unambiguously identify signatures of an outflow.  In this work, we present new millimeter interferometric observations of PG~1700 with the aim of determining whether the outflow identified in other gas phases of this source extends to the molecular gas.  

This paper is organized as follows:  In Section~\ref{sec:data} we present the new observations, describe the data reduction process, and describe several multi-wavelength archival datasets that we re-analyze for this work.  Section~\ref{sec:analysis} details the analysis of the \COIO\ data cube, the results of which are discussed in the context of existing observations in Section~\ref{sec:discussion}.  We summarize our findings in Section~\ref{sec:summary}.  Throughout this work, we adopt a cosmology of $H_0 = 73\;{\rm km\; s^{-1}\;Mpc^{-1}}$, $\Omega_{\Lambda} = 0.73$, and $\Omega_{m} = 0.27$ which gives a scale of 4.2~kpc~arcsec$^{-1}$ at $z=0.2902$.

\section{Data}
\label{sec:data}
PG~1700 is a well observed source, with data across the electromagnetic spectrum.  We combine new observations from the IRAM NOrthern Extended Millimeter Array (NOEMA) interferometer with archival {\it Hubble Space Telescope (HST)}, VLA, and Gemini Mulit-Object Spectrograph integral field spectroscopic (IFS) data to give a multi-wavelength picture of this source.  

In order to compile the multi-wavelength data, it is necessary to address differences in astrometry between various datasets.  Our approach is to apply linear shifts to center the quasar in each dataset on the coordinates of the EVN core ($\textrm{RA}=1$7\fhnd01$^{\textrm{m}}$24\fs82640, $\textrm{Dec.}=+$51\fdgnd49\arcmin20\farcs4473) identified by \citet{yang12}.  In the {\it HST} image and VLA map, we make the assumption that the quasar dominates and is centered on the pixel with the highest intensity.  In the case of the \OIII\ data, the position of the quasar is well known as the PSF center from the spatial/spectral decomposition.  For the NOEMA data, we apply a 0.48\arcsec\ primarily southern sub-pixel shift in order to place the centroid of the zero velocity map at the EVN core position.  

Registering each dataset to the quasar is a reasonable practice when the quasar obviously dominates emission, as in the VLA and optical maps.  This assumption is less certain for the NOEMA data because the PG~1700 host and companion galaxy are merging and there may be substantial \COIO\ from either galaxy.  Based on the redshifts of the \COIO\ peak and stellar absorption lines in the host and companion, it is most likely that the \COIO\ emission is centered on the quasar; the \COIO\ redshift ($z_{\textrm{CO}}=0.2899$) is within 90\kms\ of the PG~1700 host \citep[$z_{\textrm{host}}=0.2902$,][]{rupke17}, but 900\kms\ of the companion \citep[$z_{c}=0.2929$,][]{canalizo97}.  Nevertheless, in \S\ref{sec:imfit} we consider the effect of adopting the alternative scenario where we do not apply an astrometric correction, implying that the \COIO\ originates between the PG~1700 host and the companion.

\begin{figure*}[t]
\hspace{-2cm}
\centering
\includegraphics[width=0.75\textwidth]{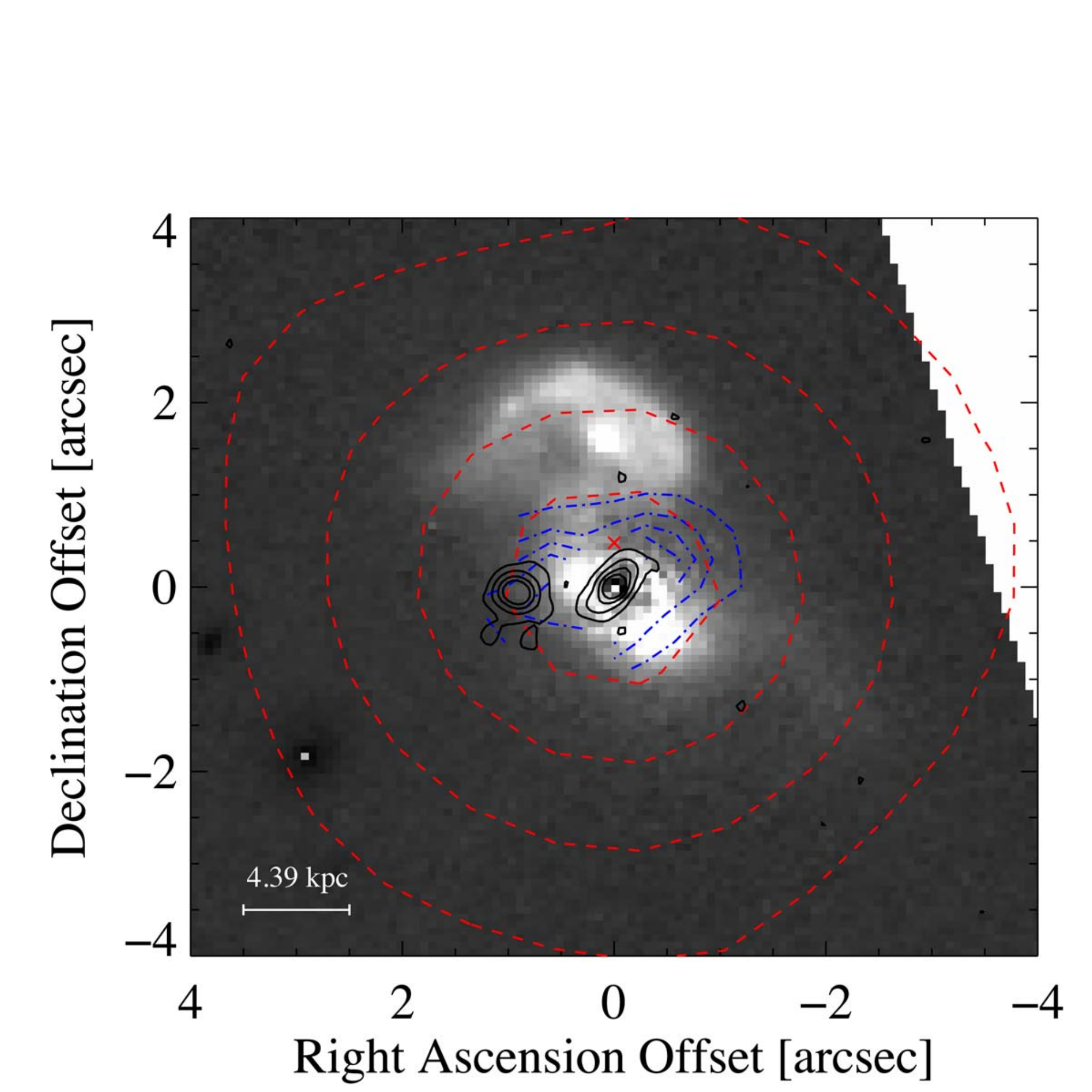}
\caption{The multi-wavelength view of PG~1700$+$518.  The grayscale shows the {\it HST} NICMOS F160W (wide H band) image of the host galaxy and its companion from \citet{evans09}.  The dashed red contours show the zero velocity map from new NOEMA observations of \COIO\ (\S~\ref{sec:mm}).  Levels are 3, 10, 20, and 30$\sigma$ where $\sigma=0.372$~mJy/beam is the rms noise in the map.  The blue dot-dashed contours map the total \OIIIw\ flux from IFS observations reported in \citealt{rupke17} (\S~\ref{sec:o3}).  Levels are 0.8, 1.7, 2.5, 3.3, and $5.0\times10^{-15}$~erg~s$^{-1}$~cm$^{-2}$~arcsec$^{-1}$.  The 8.4~GHz VLA map is shown in solid black contours (\S~\ref{sec:vla}).  Levels are 3, 10, 20, and 30$\sigma$, where $\sigma=1.798\times10^{-2}$~mJy is the rms noise in the map.  The western source has been resolved by radio interferometry into a two-sided radio jet and radio core, with the jet axis along the direction of elongation \citep{yang12}.  All of the datasets have been re-centered onto the EVN core coordinates reported by \citet{yang12} as described in \S\ref{sec:data}.  The red cross shows the position of the \COIO\ centroid if the NOEMA data are not recentered.} \label{fig:supermap}
\end{figure*}

Figure~\ref{fig:supermap} shows the {\it HST} NICMOS F160W (wide H band) image of the host galaxy and companion downloaded from the online materials provided by \citet{evans09} with contours showing VLA 8.5~GHz, NOEMA \COIO, and Gemini GMOS \OIIIw\ observations.  Below, we describe the acquisition and reduction of the individual datasets.

\subsection{New NOEMA \COIO\ observations}
\label{sec:mm}
We observed PG~1700 in CO~(1--0) with the IRAM NOEMA interferometer on 23 and 25 July 2016 (ID S16BM).  The observations were made with a total on-source time of 9.1h in the C$+$D configuration in 2 tracks, one with 8 antennas and one with 7.  Further technical details of the observations are listed in Table~\ref{tab:obslog}.  The tuning frequency was set to 89.210~GHz, the expected frequency of the redshifted CO~($1-0$) line ($\nu_{rest}=115.271$~GHz) given a redshift of $z=0.292$ based on the [O\textsc{iii}] quasar emission lines \citep{schmidt83}.  The optical narrow emission lines in PG~1700 are extremely low contrast in the early longslit spectra, and as a result redshift measurements based on the quasar spectra are inaccurate.  Detailed IFS observations of the host galaxy leverage spatial information to yield a lower redshift for the system: fitting of the integrated PG~1700 host galaxy yields $z=0.2902$ \citep{rupke17}, with the possibility of contamination due to stellar absorption from the companion.  This is the best redshift for this source and we adopt it for calculations throughout this work.

\def\th{\tablenotemark{h}}
\begin{deluxetable}{lll}
\setlength{\tabcolsep}{4pt}
\tablewidth{0in}
\tablecolumns{11}
\tablecaption{Description of observations\label{tab:obslog}}
\tablehead{
\multicolumn2c{PG~1700$+$518}
}
\startdata
RA (J2000)\tablenotemark{a} 	& 17:01:24.82640 \\
Dec (J2000)				& $+$51:49:20.4473 \\
Redshift\tablenotemark{b} 	& 0.2902 \\
Obs. date					& 2016 July 23, 25 \\
Configuration				& C, D \\
N$_{ant}$					& 8, 7 \\
Obs. freq (GHz) 			& 89.210 \\
Time on source (hr)			& 9.1 \\
Min, max baseline (m)		& 16.3--172.7 \\
FoV (arcsec)				& 56.4$\times$56.4 \\
Synth. beam (arcsec)		& 3.78$\times$3.70  \\
Synth. beam (kpc)			& 15.9$\times$15.5  \\
\enddata
\tablenotetext{a}{Coordinates of the EVN radio core component determined by \citet{yang12}.}
\tablenotetext{b}{The redshift determined by \OIIIdblt\ in the host galaxy (Rupke~et~al.\ in prep.).}
\end{deluxetable}

The data calibration was performed using the \textsc{gildas} package \textsc{clic} (version 18oct16).  At 89~GHz, the main flux calibrator was MWC349 with a flux of 1.08~Jy.  Additional flux calibrators for the July 23 track included 3C273, 1636$+$473, and 1637$+$574, where we obtained fluxes of 13.75, 0.75, and 0.72~Jy, respectively.  For the July 25 track, we obtained fluxes of 2.73 and 0.72~Jy for 1418$+$546 and 1636$+$473, respectively.  Notably, there is excellent agreement between the fluxes obtained for 1636$+$473 on the two nights of observing.  Following standard practices, the accuracy of the absolute flux calibration in the 3~mm band is better than 10\% \citep{castro-carrizo10}.  

We used the WideX correlator which has a bandwidth of 3.6~GHz and spectral resolution of 1.95~MHz, corresponding to 6.5\kms\ at the tuning frequency.  We binned the $uv$ tables by 60\kms, or approximately 10 channels.  The velocity scale is set so that the \COIO\ line peaks at zero velocity.  This requires a redshift of $z=0.2899$, which is within 90\kms\ (or 1.5 velocity bins) of the optical redshift adopted for calculations, and the redshift obtained for \COIO\ from single-dish observations \citep[$z=0.290$,][]{evans09}.  Based on visual inspection, we identified two line-free continuum regions at 84.0$-$89.07 and 89.97$-$92.00~GHz.  These were used to generate a continuum $uv$ table, which was subtracted from the WideX data to isolate a $uv$ table for \COIO.

The image cleaning and analysis was done with the \textsc{mapping} package in \textsc{gildas} (versions 26sep16 and 18-20oct).  The synthesized beam size was $3\farcs78\times3\farcs70$ at a position angle of 93\deg.  The absolute astrometric uncertainty, including both telescope and observational contributions, is of order 1/10 the beam size, or 0\farcs3.  We generated a $128\times128$ pixel map with 0\farcs8014 pixels, as recommended by the \textsc{gildas} software based on the synthesized beam size.  The image was cleaned down to 1.6 times the noise determined from an initial cleaned version of the image, using a $25\arcsec\times25\arcsec$ square support around the center of the field.  We investigated the possibility of running self calibration to improve the subtraction of residual side lobes, but typically this approach requires $S/N>20$ for success and the measured $S/N$ in the continuum before subtraction is $\sim10$ so this was ultimately not viable.  The rms noise level in the cleaned, continuum-subtracted cubes is a function of frequency, with an average flux uncertainty of 0.372~mJy across all channels.

\subsection{Archival [O\,\textsc{iii}]~$\lambda$5007 integral field spectroscopy}
\label{sec:o3}
\citet{rupke17} recently presented a new analysis of archival IFS observations of PG~1700.  We refer the reader to that work for the full details of the observations and analysis, but summarize the process here to enable a comparison between the ionized and molecular gas in the system.  PG~1700 was observed on September 23, 2003 with the integral field unit (IFU) on the Gemini Mulit-Object Spectrograph \citep[GMOS,][]{allington-smith02, hook04}.  The data were reduced using the Gemini IRAF package (v1.12), IFUDR GMOS package, and IFSRED \citep{rupke14a}.  The data reduction follows the steps outlined in \citet{rupke13b} and \citet{rupke15}, and also includes the removal of scattered light which is critical for accurate continuum and emission-line fitting.  The final output is a data cube with a 3\farcs0$\times$4\farcs2 field of view and 0\farcs3 square spaxels covering 4320--7190~\AA\ at 1.5~\AA\ resolution.

The cube was decomposed using the \textsc{IFSFIT} package \citep{rupke14b}.  This software uses \textsc{PPXF} \citep{cappellari12} to model the emission from stars in the host galaxy and \textsc{MPFIT} \citep{markwardt12} to model the quasar continuum and emission lines.  A multi-step, iterative approach makes use of both the spatial and spectral information to separate the quasar and host-galaxy contributions to the continuum and line emission.  Notably, emission from the quasar narrow-line region (NLR) in this process is handled in two ways.  First, the contribution of unresolved NLR emission is accounted for in the characterization of the quasar PSF.  After PSF subtraction, there is additional spatially resolved emission from the NLR which is of interest for studying outflows and is therefore analyzed further.  These resolved emission lines are modeled in each spaxel with 0--2 Gaussian components, where individual components are included only when they are statistically significant at greater than the 3$\sigma$ level.  The Gaussians are then sorted into two maps, such that the priority is to facilitate a smooth galaxy rotation curve in the first map.  If an outflow is present, it will show up in the second map (although the second map is not necessarily an outflow).  

\begin{figure*}
\begin{minipage}[!b]{0.33\textwidth}
\centering
\includegraphics[width=6.5cm]{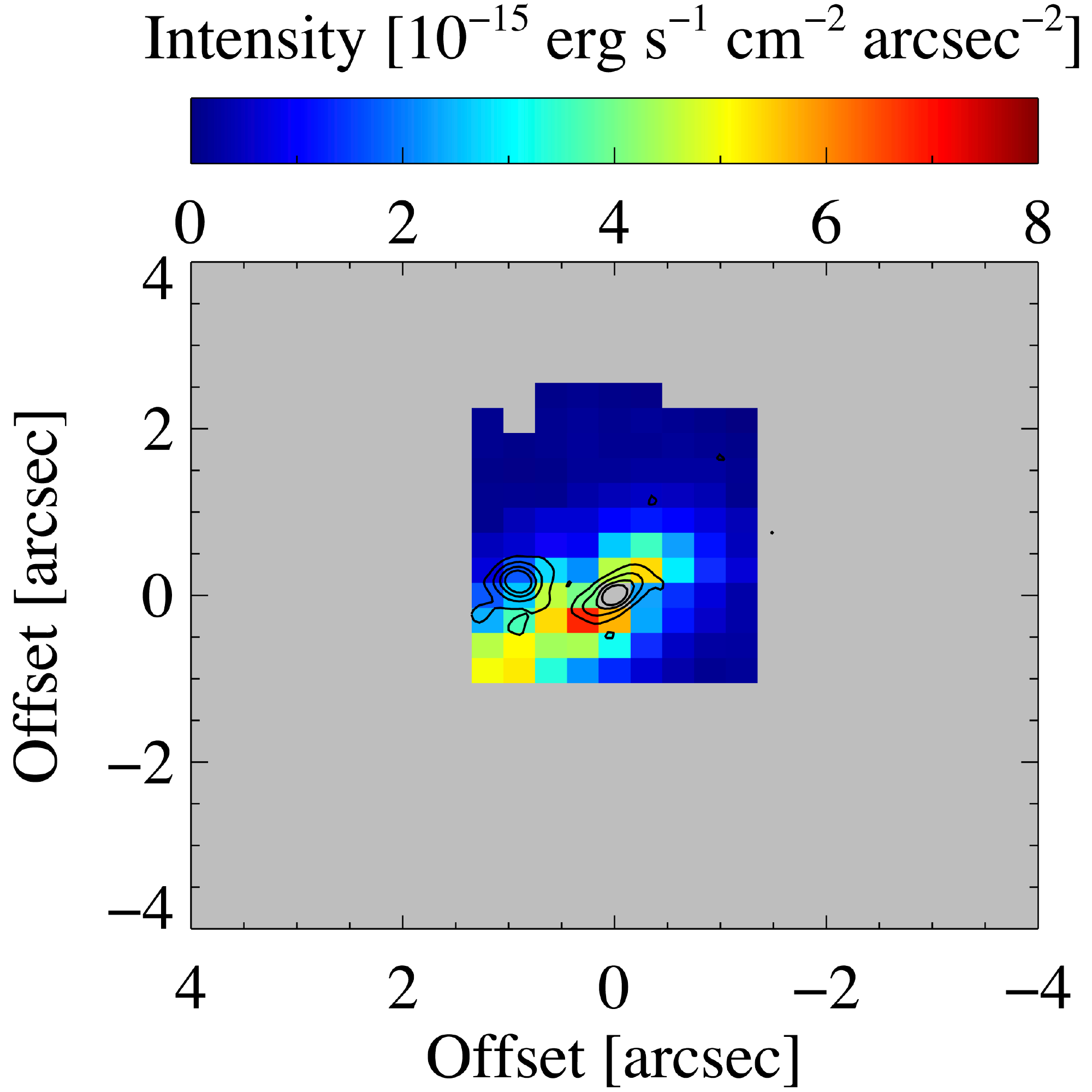}
\end{minipage}
\hspace{0.25cm}
\begin{minipage}[!b]{0.33\textwidth}
\centering
\includegraphics[width=6.5cm]{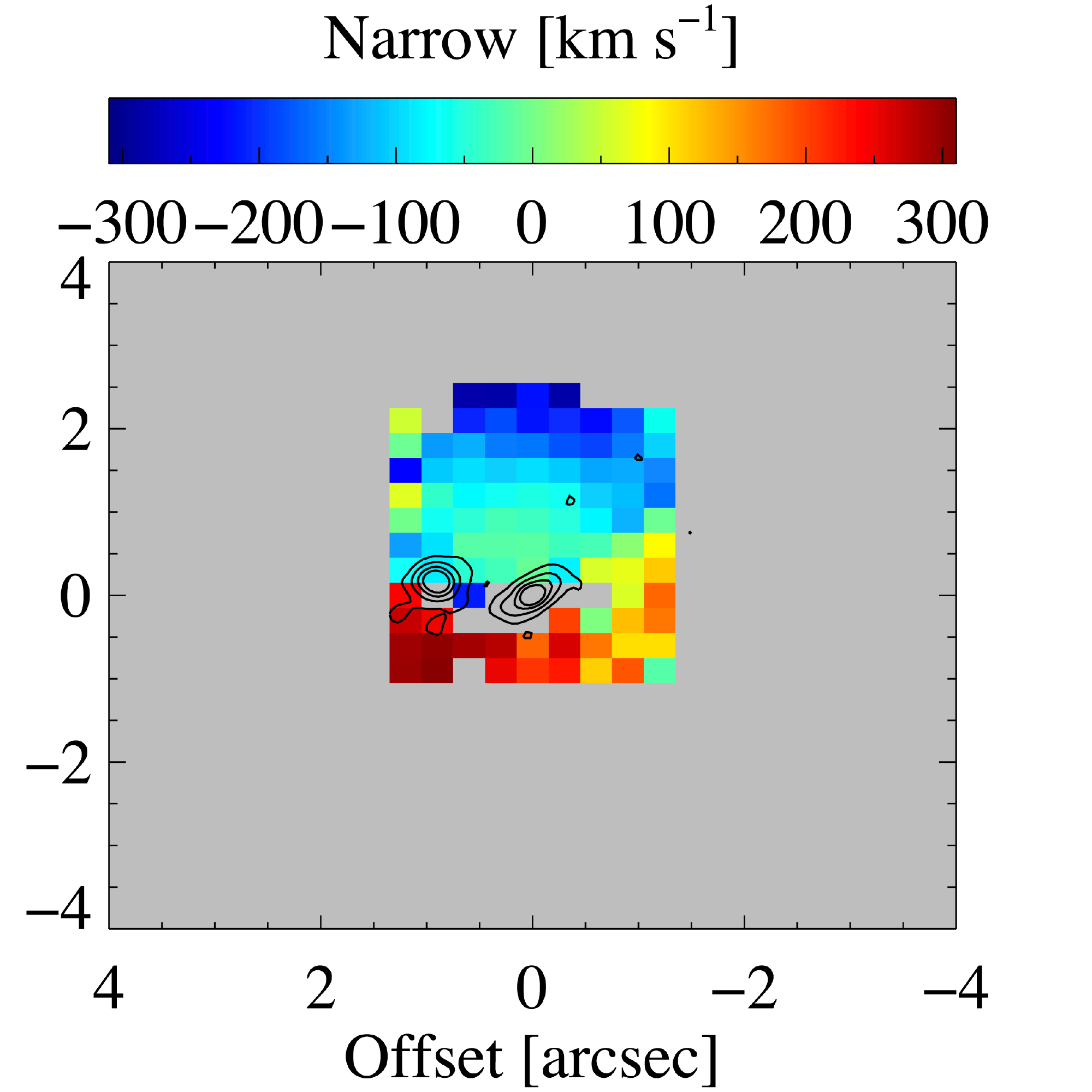}
\end{minipage}
\hspace{-0.25cm}
\begin{minipage}[!b]{0.33\textwidth}
\centering
\includegraphics[width=6.5cm]{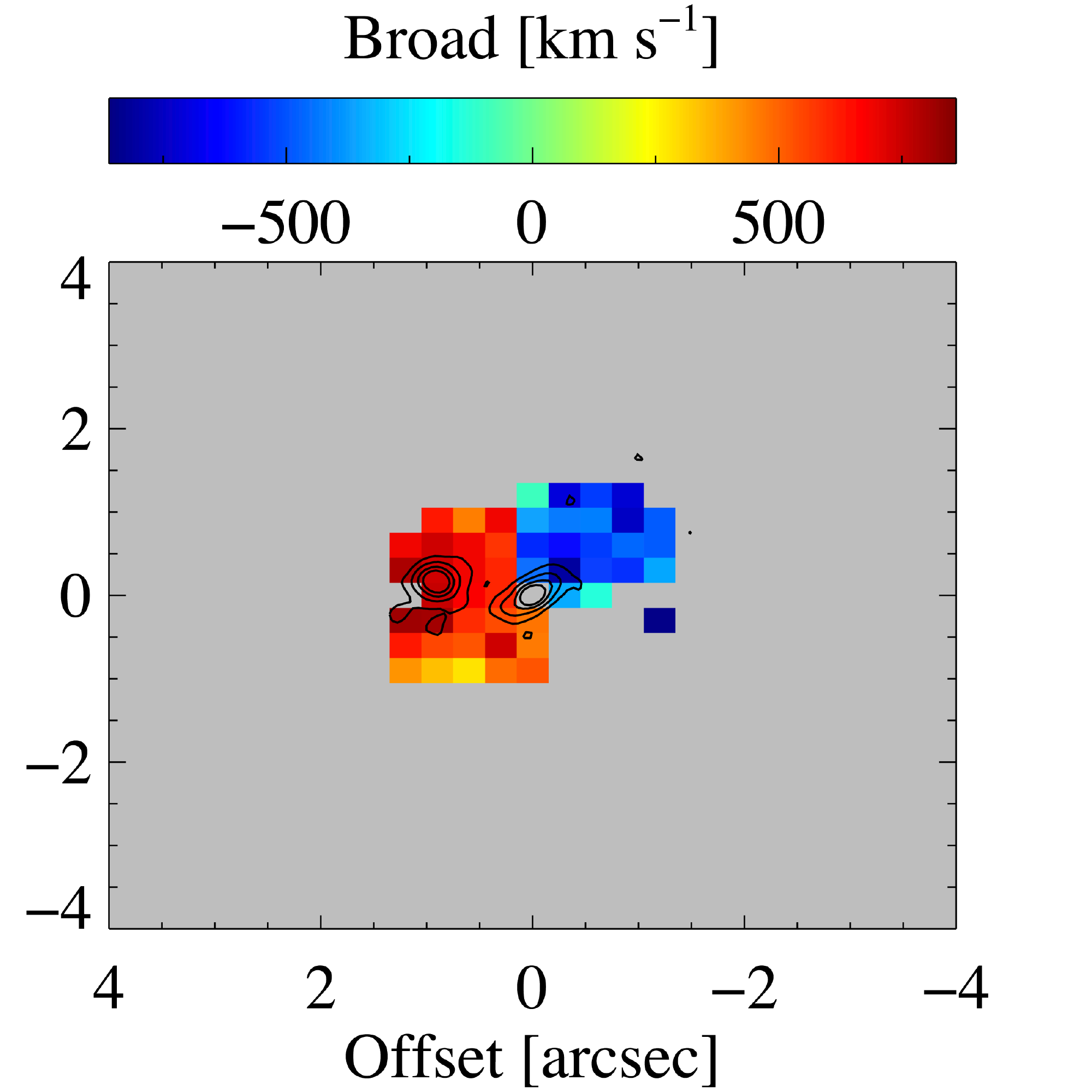}
\end{minipage}
\caption{Maps of the spatially resolved \OIIIw\ emission in PG~1700.  Panels show the total integrated flux, peak offset of the narrow velocity component used to model the \OIII\ emission-line profile, and the peak offset of broad velocity component that traces a jet-driven outflow.  The black contours show the VLA radio map, with the jet axis along the elongated source near the origin.  Up is 14~degrees east of north.  See \S\ref{sec:o3} for details of the spectral analysis used to derive these measurements.  The physical origin of the narrow \OIII\ emission is unclear; it may be N/S rotation misaligned with the stellar bulge \citep{evans09,veilleux09}, tidal interactions with the companion, interactions with the jet, or some combination of these.  However, the \OIII\ is most intense and is driven to high velocities along the direction of the radio jet.  \label{fig:o33map}}
\end{figure*}

Figure~\ref{fig:o33map} shows the maps of the resolved \OIII\ emission in PG~1700.  The total \OIII\ flux is concentrated along the axis of the radio jet identified by \citet{yang12}.  The velocity map of the narrow \OIII\ component (the first in the previous paragraph) does not have a clear physical interpretation: it may trace rotation in a disk, tidal interactions with the companion galaxy, interactions with the compact radio jet, or a combination of these.  If this map traces rotation, it implies that the low-velocity \OIII\ emission, which is aligned primarily along the N/S direction, is misaligned with the NE/SW position angle identified for the stellar bulge \citep{evans09,veilleux09}.  There is a jet-driven outflow seen in \OIII\ that is apparent in the velocity map of the second, broad component.

\subsection{Archival VLA 4.86 and 8.5~GHz observations}
\label{sec:vla}
PG~1700 was observed with the VLA at 4.86 and 8.44~GHz with a bandwidth of 50~MHz on 15 December 1988 in A configuration (project code AB0512).  We obtained the raw datasets from the VLA archive, including calibration scans.  The flux calibrators observed for this project were 1739$+$522 at 4.86~GHz, at an assumed 1.2 Jy flux density and 3C286 at 8.5~GHz.  At both frequencies, 1739$+$522 was used for phase calibration. The data flagging and reduction followed standard techniques using CASA version 4.0.1.  The full image at each frequency was processed individually with the CLEAN algorithm, with a maximum of 5000 iterations, a gain of 0.1, and natural weighting.  The final maps have rms noise values of 2.41$\times10^{-5}$~Jy and 1.798$\times10^{-5}$~Jy at 4.9 and 8.5~GHz, respectively.  The absolute flux calibration uncertainty is 10\% \citep{baars77,perley17}, and dominates other sources of uncertainty.      

The final maps at both frequencies show two sources separated by approximately 1\arcsec.  Both sources are detected at high signal-to-noise ratio ($>5\sigma$) at both frequencies.  High-resolution radio interferometry reveals that the western source is a radio core with nearly symmetric jets directed along the axis of elongation in the VLA maps \citep{yang12}.  The nature of the eastern source is not completely clear (see \S~\ref{sec:discussion} for further discussion). Using CASA, we modeled both VLA sources with elliptical Gaussians in order to determine their properties and measure radio spectral index.  The best-fitting parameters are listed in Table~\ref{tab:vla}.  Adopting the convention $S_{\nu}\propto \nu^{-\alpha_{r}}$, we find radio spectral indices between 4.86 and 8.44~GHz of 0.45 and 0.52 for the eastern and western (core) sources, respectively.  This tendency for the eastern source to have a slightly flatter radio spectrum was noted by \citet{kukula98}.

\def\th{\tablenotemark{h}}
\begin{deluxetable}{lll}
\tablewidth{0in}
\tabletypesize{\scriptsize}
\tablecolumns{3}
\tablecaption{Measured radio properties from VLA observations\label{tab:vla}}
\tablehead{
\colhead{} &
\colhead{VLA E} &
\colhead{VLA W} 
}
\startdata
{\bf 4.86~GHz}				&								& \\
RA (J2000)				& 17:01:24.91407 					& 17:01:24.81518 \\
Dec (J2000)				& $+$51:49:20.3833					& $+$51:49:20.4463 \\
Gaussian size				&  0\farcs459$\times$0\farcs390		& 0\farcs532$\times$0\farcs386 \\
Position angle (deg)			& 15.6							& 153.907 \\
Integrated flux\tablenotemark{a} (mJy)			& 2.57							& 1.98 \\
Peak flux\tablenotemark{a} (mJy beam$^{-1}$) 	& 1.98							& 1.32 \\
\\
{\bf 8.4~GHz}				&								& \\
RA (J2000)				& 17:01:24.91308 					& 17:01:24.81518\\
Dec (J2000)				& $+$51:49:20.3832					& $+$51:49:20.4463\\
Gaussian size				&  0\farcs302$\times$0\farcs285		& 0\farcs398$\times$0\farcs217 \\
Position angle (deg)			& 24.4						& 147.224 \\
Integrated flux\tablenotemark{a} (mJy)			& 1.55							& 1.49 \\
Peak flux\tablenotemark{a} (mJy beam$^{-1}$) 	& 0.89							& 0.86 \\
\\
$\alpha_{r}$				& 0.45							& 0.52 
s
\enddata
\tablenotetext{a}{Uncertainties are dominated by the absolute flux uncertainty, which is 10\%.}
\end{deluxetable}

\section{Analysis of NOEMA \COIO\ data}
\label{sec:analysis}

\subsection{Spectral decomposition and physical properties}
\label{sec:specfit}
\begin{figure}[t]
%
\hspace{-0.77cm}
\includegraphics[width=0.53\textwidth]{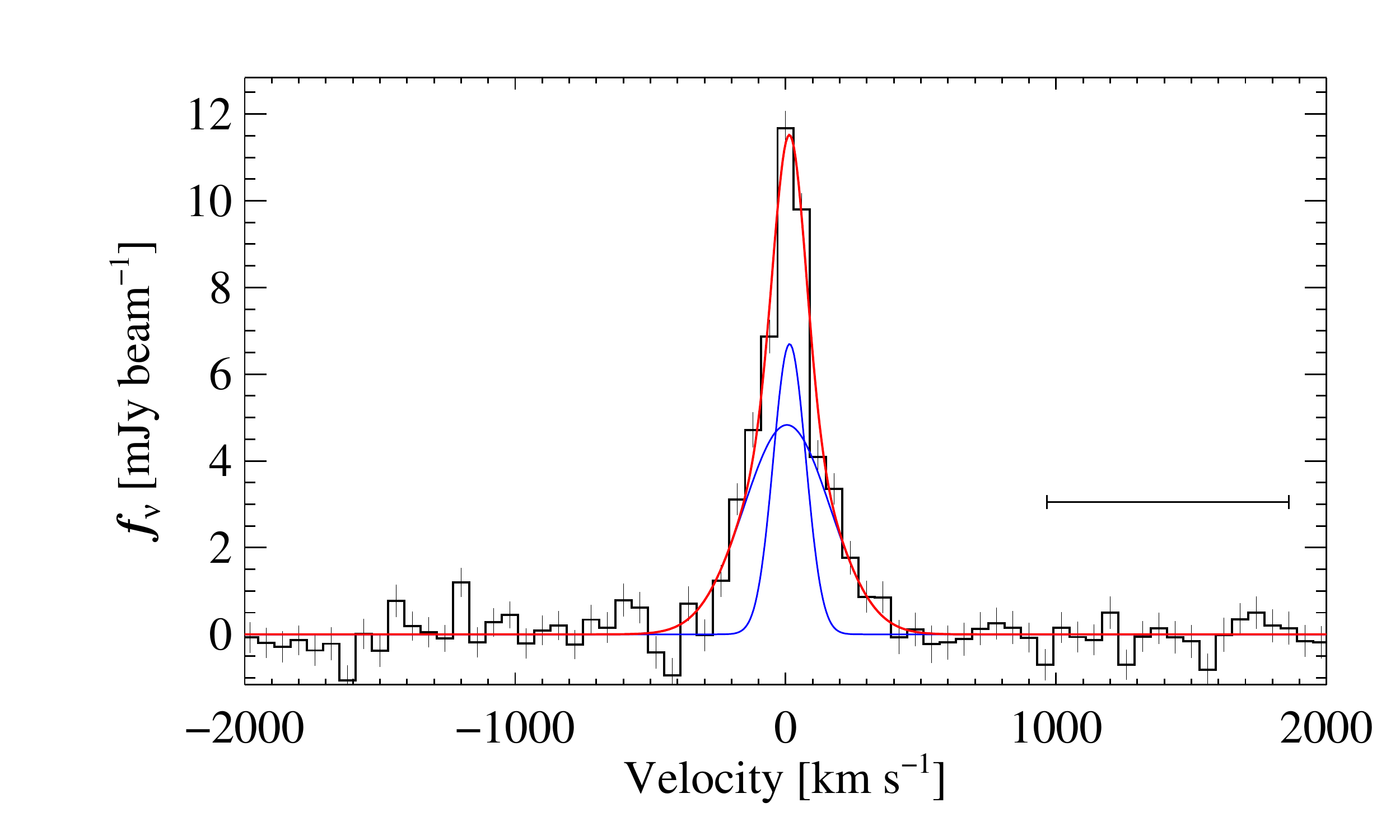}
%
\caption{The spectral decomposition of the continuum-subtracted \COIO\ line profile in the central spaxel.  Our procedure provides a good fit to the line, which is above the 1$\sigma$ level out to velocities of $\pm300$\kms.  The total model is shown in red, while the two Gaussian components used in the fit are in blue.  The horizontal bar shows the level of 5 times the noise in the continuum required for a robust detection of the line and its extent identifies the part of the spectrum where the noise was measured.  } \label{fig:specfit}
\end{figure}

In order to measure the properties of the \COIO\ line profile and investigate their variation with spatial position, we decomposed the continuum-subtracted spectra.  We modeled the profile with two Gaussian components, which were allowed to vary independently and to which no independent physical meaning is attributed.  To facilitate meaningful results, we required that the fluxes of both Gaussians be positive, the widths be larger than 60\kms, and the position of the second Gaussian be within $+340$/$-110$\kms\ of the position of the nominal CO wavelength.  Collectively, these constraints prevent one or both Gaussians from being used to model noise spikes in the spectrum in the automated fit and we visually inspect each fit to ensure that this holds true.  The best-fitting model was then determined by minimizing the chi-squared statistic using the IDL package \textsc{mpfit} \citep{markwardt12}, which implements a Levenberg-Marquardt least-squares method, with all channels weighted by their measured noise.  An example of the spectral decomposition for the central spaxel is shown in Figure~\ref{fig:specfit}.

In spaxels where the \COIO\ line was robustly detected, we measured the properties of the line profiles.  We quantified robust detections of the line profile as cases where the peak of the best-fitting model is at least 5 times the noise in the spectrum, where the noise was taken to be the rms of the continuum in the range 1000$-$2000\kms.  We measured the integrated flux, the velocity centroid, and the velocity dispersion of the line profiles.  All measurements were made on best-fitting model of the spectrum (as opposed to the data themselves) within 2000\kms\ of the rest frame.  We carried out simulations in order to determine the uncertainties on measured properties of the spectra.  In these simulations, we repeated the spectral line measurements on synthetic spectra generated by perturbing the flux in each channel according to a Gaussian distribution with width equal to that channel's noise.  After 1,000 trials we verified that the resulting distribution for each measured property had a mean comparable to our measurement and took the standard deviation of the distribution as our uncertainty in that measurement.  The formal fit uncertainties are in the range $0.03-0.12$~Jy~\kms~beam$^{-1}$, $2-50$\kms, and $2-130$\kms\ for the integrated flux, velocity centroid, and velocity dispersion, respectively.  Adding the velocity resolution of the original spectra in quadrature yields a minimum velocity uncertainty of 6\kms.  We note that these formal uncertainties resulting from the spectral decomposition are often substantially smaller than those resulting from differences between adopting different fitting and measurement procedures \citep[for a discussion, see][]{runnoe15a}.

\begin{figure*}
\begin{minipage}[!b]{0.33\textwidth}
\centering
\includegraphics[width=6.5cm]{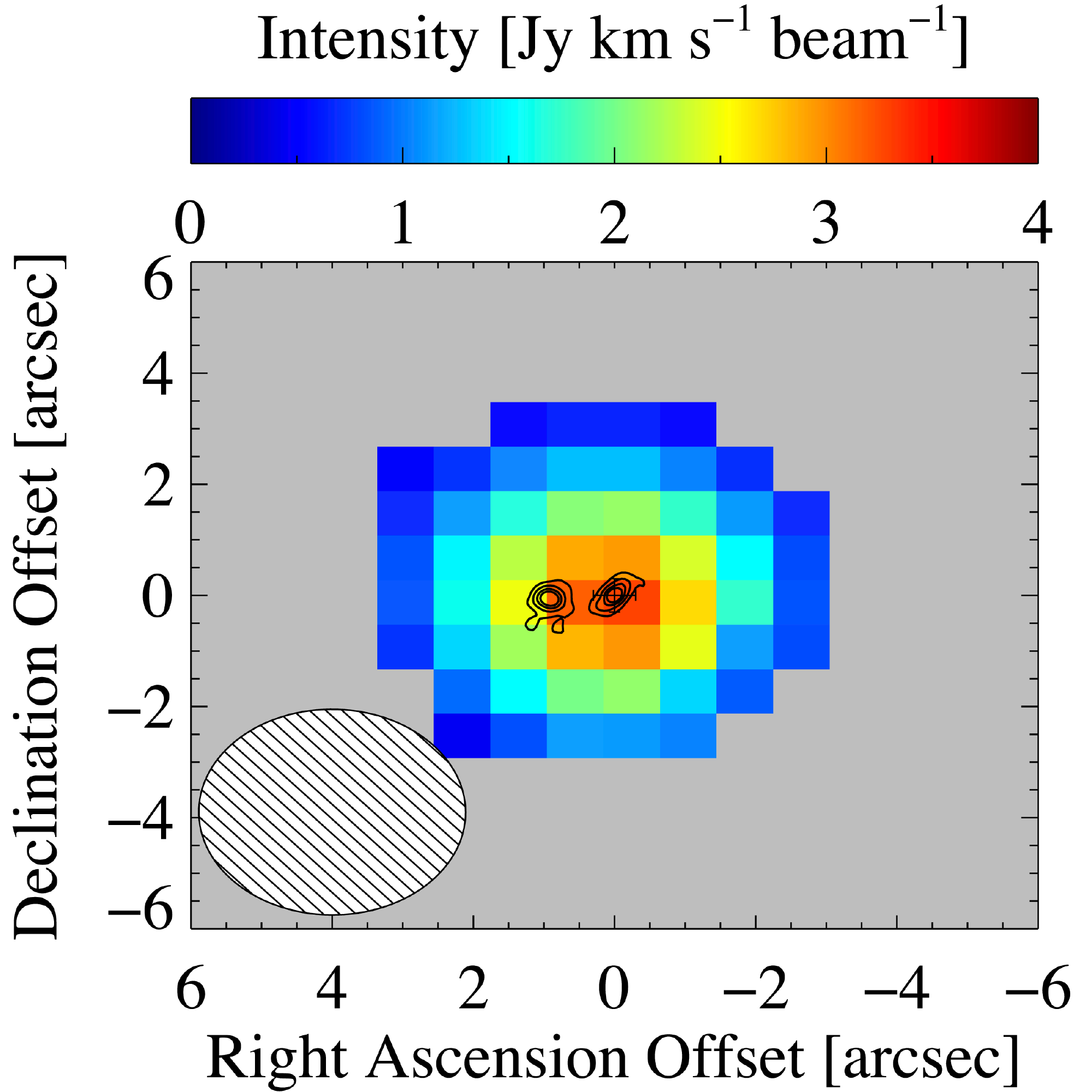}
\end{minipage}
\hspace{0.25cm}
\begin{minipage}[!b]{0.33\textwidth}
\centering
\hspace{-0.1cm}
\includegraphics[width=6.5cm]{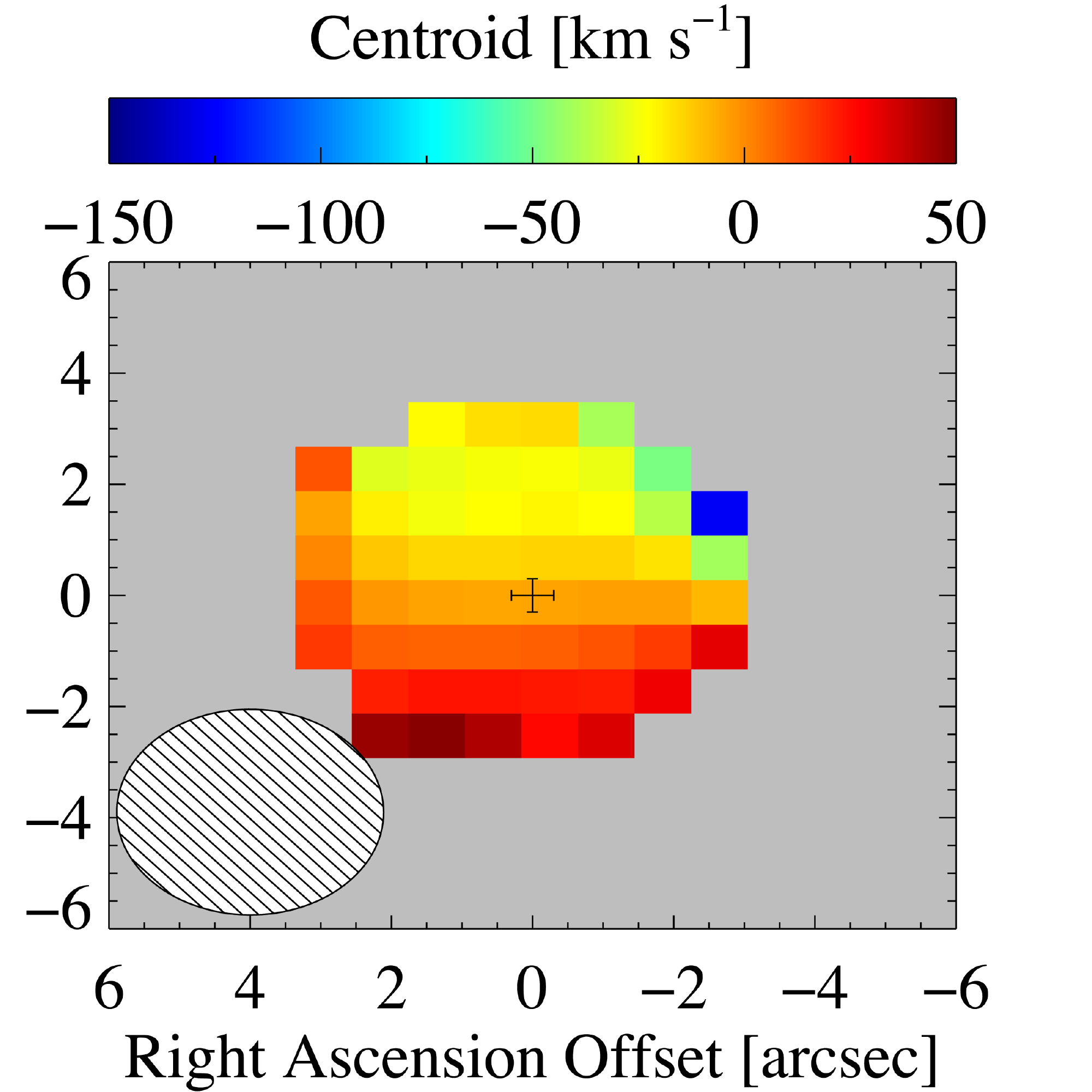}
\end{minipage}
\hspace{-0.1cm}
\begin{minipage}[!b]{0.33\textwidth}
\centering
\includegraphics[width=6.5cm]{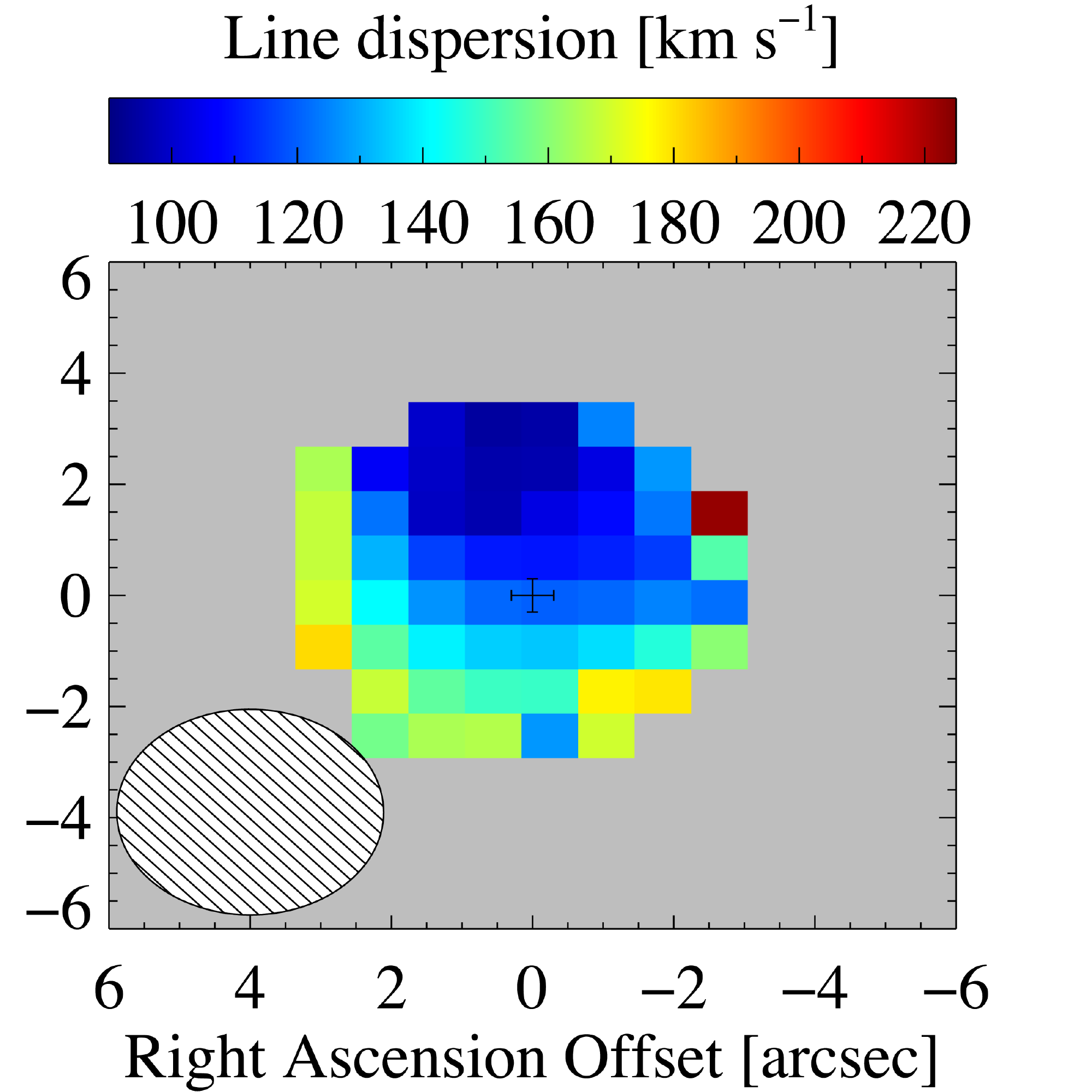}
\end{minipage}
\caption{Maps of the integrated flux, velocity centroid, and velocity dispersion of the \COIO\ emission line for PG~1700.  The black contours in the first panel show the VLA radio map, with the jet axis along the elongated source near the origin.  The systematic drift in centroid and smooth trend in dispersion may be a subtle indications of interaction between the radio jet and molecular gas.  Formal intensity uncertainties are $0.03-0.12$~Jy~beam$^{-1}$,  centroid uncertainties are $2-50$\kms, and dispersion uncertainties are $2-130$\kms.  These maps include measurements for spaxels where the total line profile was detected at the 5$\sigma$ level, as described in Section~\ref{sec:specfit}.  The black cross shows the location of the EVN core \citep{yang12} with uncertainty given by propagating the uncertainties on the NOEMA and EVN absolute positions.  \label{fig:3map}}
\end{figure*}

The spatial maps of the measured spectral properties are shown for the total line profiles in Figure~\ref{fig:3map}.  We set the zero point for spatial offsets in these maps using the position of the radio core in the EVN core from \citet{yang12}, as described in \S\ref{sec:data}.  Uncertainties on the zero offset are determined by propagating the uncertainties in the absolute NOEMA and EVN positions, and are dominated by the NOEMA uncertainties.  The intensity map is generally similar to a point source, but the centroid and line dispersion maps show systematic trends.  The velocity centroid drifts by approximately 100\kms\ over the extent of the source.  While this size corresponds to only 1--2 times the beam size, given uncertainties of $<50$\kms\ this may be a real effect.  It is more difficult to determine whether the trend in line dispersion is real.  The uncertainties can be larger than the size of the effect seen in the maps, and it is also hard to tell to what extent S/N plays a role in producing it.  On one hand, the spaxels with the broadest lines are those near the edge with the lowest S/N, while on the other the effect of low S/N is to systematically decrease line dispersion measurements \citep[e.g.][]{denney09a}.  However, the trend is smooth and a visual inspection of the spectra in spaxels with the broadest line dispersion measurements does suggest that they are broader than those nearer the center of the source, although the high-velocity gas is typically not detected above our 5$\sigma$ threshold.  

We also derived physical properties based on the spectral decomposition of the \COIO\ emission line.  First, we calculate the CO luminosity, $L_{CO}^\prime$.  We measure a total CO flux of $S_{CO}\Delta v = 3.12\pm0.02$~Jy~km~s$^{-1}$ and obtain $L_{CO}^\prime = (1.208\pm0.009)\times10^{10}$~K~km~s$^{-1}$~pc$^{2}$ given $z(=0.2902)$ and $D_{L}(=1442.2\textrm{ Mpc})$.  Following \citet{evans09} who performed the calculation based on single-dish observations, we calculate a molecular gas mass, $M(\textrm{H}2)$, of $(4.83\pm0.04)\times10^{10}$~\Msun\ assuming the conversion factor for an unresolved source $\alpha = M(\textrm{H}2)/L_{CO}^\prime \sim 4 \textrm{\Msun} (\textrm{K km s}^{-1}\textrm{pc}^2)^{-1}$.  

\subsection{Spatial image analysis}
\label{sec:imfit}
\begin{figure}
\hspace{-0.32cm}
\includegraphics[width=0.511\textwidth]{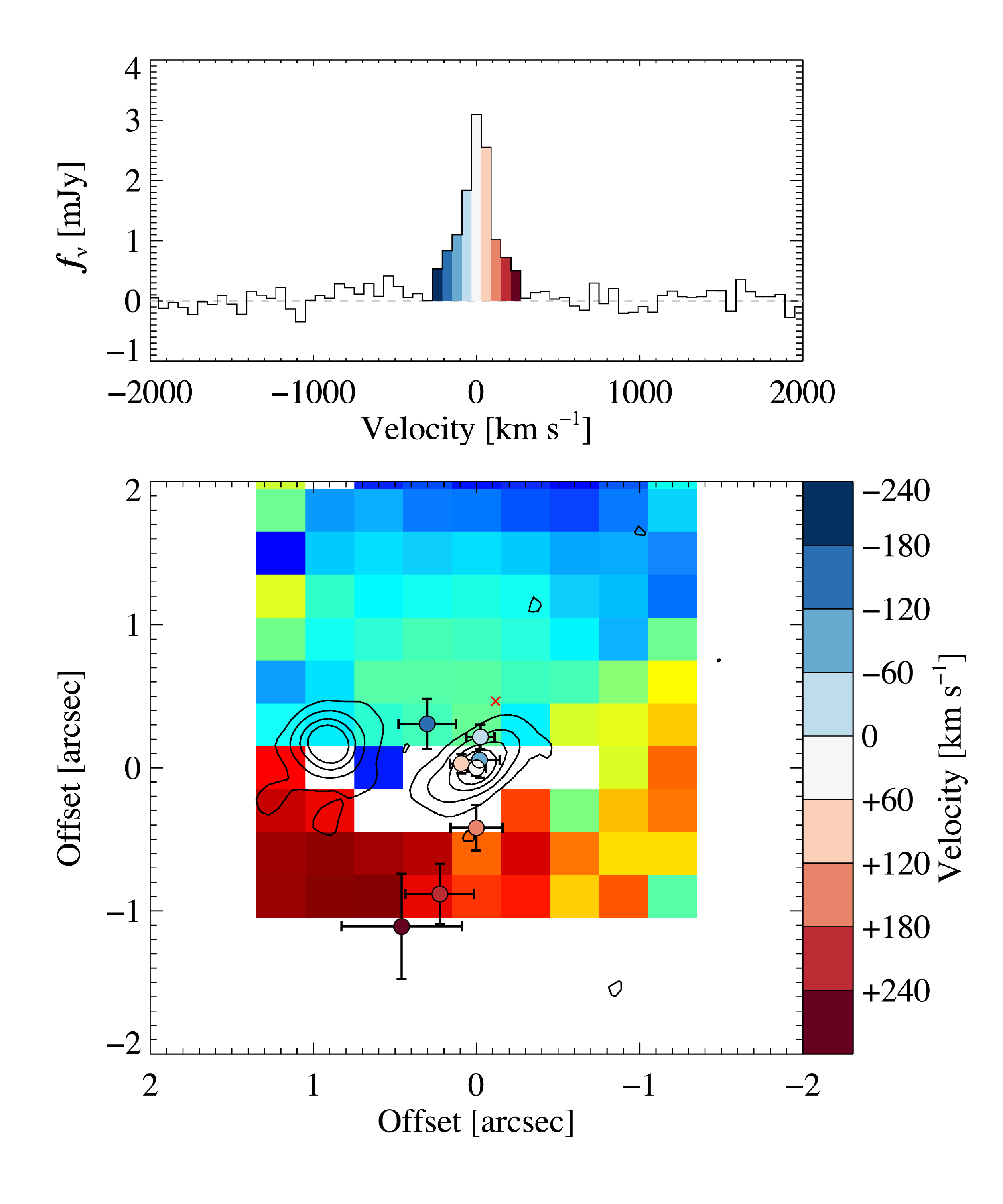}
\caption{Top: the total \COIO\ spectrum integrated within an $8\times8$\arcsec\ box.  Color coded bins identify individual velocity slices through the CO data cube where the source was detected at greater than 5$\sigma$ significance.  Bottom: the spatial position of the centroid determined from a point-source fit to the visibilities in the {\it uv} plane, as described in Section~\ref{sec:imfit}.  The channel map for the most blueshifted bin is dominated by noise, so that measurement is not shown.  The frame has been rotated to match the IFS observations, where up is 14~degrees east of north and the IFS color scheme is identical to the center panel of Figure~\ref{fig:o33map}.  The black contours in the first panel show the VLA radio map, with the jet axis along the elongated source near the origin.  The red cross shows the position of the \COIO\ centroid if the NOEMA data are not recentered on the EVN core (\S\ref{sec:data} and \ref{sec:imfit}).  The \COIO\ centroids generally follow the low-velocity \OIII\ map, shown in the background map.  Notably, the \COIO\ centroids in the two most redshifted bins match the axis of the radio jet and the jet-driven \OIII\ outflow.  This agreement is strengthened if we do not recenter the NOEMA map on the EVN core.} \label{fig:centroid}
\end{figure}

In order to better characterize the velocity centroid drift seen in the \COIO\ spectral maps, we can determine the image centroid as a function of velocity, which is well defined.  Because the source is not complex in the image plane, we can make this measurement in the {\it uv} plane, directly from the visibilities, rather than doing a two-dimensional image decomposition (although we performed this exercise and obtained qualitatively similar results).  This has the advantage of being independent of the uncertainty associated with cleaning the image during the data reduction.  Instead, the largest uncertainty is associated with the selection of a model, where there is little ambiguity in this case.  We used the \textsc{uvfit} task in Gildas and adopted a point source model.  The \textsc{uvfit} routine fits the visibilities directly and produces measurements of the image centroids with uncertainties.  In Figure~\ref{fig:centroid} we compare the centroid measurements as a function of velocity with the multi-wavelength data.  The \COIO\ centroids align well with the signal seen in the narrow \OIII\ IFS measurements, and also with the radio jet at the most redshifted velocities (where both the broad and narrow components follow the jet axis in the \OIII\ data).    

In \S\ref{sec:data} we described the application of linear spatial shifts that we adopted in order to center all of the multi-wavelength datasets on the quasar.  This assumption is reasonable for maps where the quasar clearly dominates the emission and can be used for registration, but it is less clear whether it is fair for the \COIO\ map because PG~1700 is a merging system.  In Figure~\ref{fig:centroid} we show a red cross to mark the unshifted center of the zero velocity \COIO\ emission.  If we adopt the nominal astrometry for the NOEMA dataset,  \COIO\ centroids would shift so that the off-white zero-velocity point falls at this position.  This has the effect of shifting the centroids for the reddest \COIO\ bins into better alignment with the most redshifted \OIII, strengthening the agreement between the \COIO\ centroid alignment, the jet, and the IFS \OIII\ outflow.  Thus, our default practice of registering all the datasets is the most conservative approach.

\section{Discussion}
\label{sec:discussion}
The new millimeter observations are fully consistent with the existing single-dish measurements from \citet{evans09}.  The \COIO\ line shape, and thus velocity distribution of the gas, is similar and the redshifts match within the 60\kms\ spectral resolution.  The derived physical properties are comparable as well, once differences in cosmology and analysis (which are not reflected in the formal uncertainties) are accounted for.  Thus, the main improvement provided here is the spatial resolution, which is nearly a factor of ten better than the previously available 26\arcsec.  That said, with a beam size of $3.8$\arcsec, the NOEMA observations of PG~1700 may suffer contamination from emission in the companion galaxy located 1\arcsec\ away.  Nevertheless, as \citet{evans09} point out, the \COIO\ redshift matches PG~1700 and not the companion \citep[$z=0.2929$,][]{canalizo97}.  The companion redshift is measured reliably by those authors from IFS observations of stellar absorption lines, and the difference in redshift between the \COIO\ and the companion corresponds to approximately 900\kms.  This suggests that it is the primary source of \COIO\ emission in the system.

By considering the new \COIO\ observations in the context of existing multi-wavelength data, we can better appreciate the full picture of the effect of the radio jet in PG~1700.  In the VLA map, there are two radio sources.  The nature of the eastern VLA source is not completely clear.  It is resolved out in the EVN map, making a background AGN unlikely.  The radio spectral index measurement is consistent with synchrotron emission from an aged electron population \citep{feretti07}, indicating a relic radio source.  Two likely explanations are that the jet either re-oriented or was redirected.  In the first case, the jet may have pointed from the radio core in the direction of the eastern VLA source.  It quickly re-oriented, possibly due to interaction with the companion galaxy, and the emission at the position of the eastern VLA sources has not yet faded.  Alternatively, the eastern VLA source is the result of the VLA jet being redirected by material in the host galaxy.  This explanation, suggested by \citep{yang12}, is supported by the direction of the \OIII\ outflow, which is along the current jet axis but has the most redshifted component in to the southeast right at the position of the eastern VLA source.  Additionally, \citet{stockton98} find a spot of intense star formation at the end of the radio jet, in support of this interpretation.  Finally, the simulations of \citet{mukherjee16} show this behavior, where the morphology of the radio plasma has an irregular morphology as it moves through the cavities in the clumpy ISM.  

The western VLA source, which includes the radio core and nearly symmetric radio jet \citep{yang12}, is driving an ionized outflow seen in the IFS \OIII\ observations \citep{rupke17}.  We do not see strong evidence for an outflow of molecular gas on the size scales probed by our observations, of order several arcseconds corresponding to $\sim$15~kpc at $z=0.2902$.  However, the velocity centroid of the \COIO\ line has a slight (of order 100~km~s$^{-1}$/kpc, see Figure~\ref{fig:3map}) drift in the north/south direction, matching the low-velocity \OIII\ in the host galaxy.  As shown in Figure~\ref{fig:centroid}, in the most redshifted velocity bins the spatial centroid of the two-dimensional image drifts along the jet axis.  At the positions corresponding to the extremely red centroids, the CO line has systematically broader width than in other positions (see Figure~\ref{fig:3map}).  It is not immediately clear how reliable these extreme width measurements are: they correspond to the lowest S/N spaxels, but do appear broader based on visual inspection.  Also, compared to the ionized gas traced by \OIII, the molecular gas has much lower velocity, similar to the low-velocity core of the \OIII\ line.  If the jet interacts with both the ionized and molecular gas phases, it will naturally drive the ionized gas to higher velocities.  

These pieces of evidence may be a hint that the jet does affect the molecular gas on smaller scales.  This would be consistent with a scenario where the high-powered jet ($P_{jet}\sim10^{45-46}$~erg~s$^{-1}$) is accelerating dense gas clumps along the shocked surface of an energy bubble, to velocities of hundreds of \kms.  This matches the velocities reached in the simulations of \citet{mukherjee16}, although the gas conditions do not reach the temperatures/densities appropriate for cold molecular gas.  However, they do demonstrate that lower density gas is accelerated to lower velocities in the outflow.  In the simulations of \citet{mukherjee16}, these high-power jets that have the ability to effectively drive outflows, although they are less able to affect star formation over a large volume in their hosts as they pierce through the ISM.  The NOEMA beam size is relatively large compared to the size of the jet, which may prevent us from clearly detecting any signs of interaction.  Millimeter interferometric observations with higher spatial resolution and sensitivity would be able to distinguish this scenario by tracing spatial distribution of the low-contrast, high-velocity wings of the CO line.

\section{Summary}
\label{sec:summary}
In this work, we present new NOEMA millimeter interferometric observations of the \COIO\ line in the type 1 loBAL quasar PG~1700+518.  The observations were made in the C+D configuration, yielding a beam size of $3\farcs78\times3\farcs70$ and a map with 0\farcs8014 pixels.  This is over half an order of magnitude improvement in spatial resolution over existing single-dish observations.  Following a spectral decomposition, we made maps of the \COIO\ intensity, velocity centroid, and velocity dispersion.  These show a $\sim$100\kms\ drift in the north/south direction in velocity centroid and a trend of approximately the same magnitude in the southeast/northwest direction in the line dispersion over a physical region corresponding to 1--2 times the beam size.  We also fit the {\it uv} visibilities with a point-source model in order to trace the centroid of gas in 60\kms\ velocity bins.  The \COIO\ centroids match the low-velocity \OIII\ seen in the IFS observations, albeit at lower velocities consistent with the core of the \OIII\ line.  The physical origin of this low-velocity \OIII\ emission is not completely clear; it may be rotation in the host galaxy (albeit misaligned with the stellar bulge), tidal interactions with the companion, interactions with the jet, or a combination of these.  At the highest velocities, the \OIII\ traces a jet-driven outflow.  Similarly, in the two most redshifted bins the \COIO\ centroids fall along the axis of the compact radio jet, which may hint that the jet interacts with the molecular gas on smaller spatial scales.  Combining these observations with results from existing multi-wavelength data we obtain the following:
\begin{itemize}
\item With the improved spatial resolution of NOEMA over previous single-dish observations, we are at the cusp of spatially resolving the CO emission from PG~1700.  We cannot fully separate contributions to the \COIO\ emission that may come from the PG~1700 host galaxy and its companion $\sim$1\arcsec\ away.  However, the CO and host redshifts match, suggesting that it is the primary CO source in the system.  Physical properties measured from the \COIO\ emission line are consistent with previous single-dish measurements.
\item There are no completely unambiguous signs of a high-velocity ($v>100$\kms) outflow in the molecular gas on the arcsecond scales of our observations.  This is in contrast to the ionized gas traced by \OIII, which is clearly driven to high velocities by the radio jet.  The tendency for the gas in the red wing of the \COIO\ emission line to be broader and to be distributed in the direction of the jet axis may be a hint that the jet does affect the molecular gas.  In this case, it may be that the signs of interaction are present on small scales and the new NOEMA observations do not have sufficiently high spatial resolution to discern them.
\item  We interpret the multi-wavelength picture of PG~1700 by drawing on the results of \citet{mukherjee16}.  In their simulations, the high-power radio jets expand an energy bubble which efficiently drives radial outflows along its surface.  The jet potentially adopts an irregular morphology as it searches for the path of least resistance through the dense filaments of gas.  These jets have enough momentum to punch through the ISM relatively quickly, so they are less able to affect the star formation properties of their hosts over a large volume. Observations of PG~1700 are consistent with such a picture.  The compact radio jet in PG~1700 has high power ($P_{jet}\sim10^{45-46}$~erg~s$^{-1}$), similar to those in the simulations.  The compact radio jet drives a high-velocity \OIII\ outflow, with the most redshifted ionized gas coincident with the eastern VLA source.  Finally, the most redshifted \COIO\ hints at small-scale interactions between the jet and dense molecular gas, but higher spatial resolution observations will be required to confirm this.  The alternative possibility is that the \COIO\ motions are driven by rotation in a galaxy disk or tidal interactions with the companion galaxy.
\end{itemize} 
In the context of observing jet-driven molecular outflows, additional observations will have a lot of deciding power in this system.  Although it does have a jet, PG~1700 is formally radio quiet and the jet is very compact.  Thus, higher spatial resolution observations with the sensitivity to map the low-contrast, high-velocity wings of the \COIO\ emission line would be effective at determining whether the hints seen here are signs of a jet-driven outflow on smaller scales.  

\acknowledgements

JCR acknowledges the extraordinary support of Charl\`{e}ne Lef\`{e}vre in reducing the NOEMA data, as well as the hospitality of IRAM where some of this work was carried out.  JCR and KG thank Edmund Hodges-Kluck for helpful discussions during the preparation of this work.  D.S.N.R. was supported in part by the J. Lester Crain Chair of Physics at Rhodes College and by a Distinguished Visitor grant from the Research School of Astronomy \&\ Astrophysics at Australian National University.  The authors thank the anonymous referee for suggestions that improved the clarity of this work.  This work was based on observations carried out under project number S16BM with the IRAM NOEMA interferometer. IRAM is supported by INSU/CNRS (France), MPG (Germany) and IGN (Spain).


\bibliographystyle{apj}
\bibliography{all.102416}
\clearpage

\label{lastpage}
\end{document}